\documentclass[11pt]{article}
\usepackage{moriond,epsfig}

\def\noi{\noindent}

\def\bea{\begin{eqnarray}}  \def\eea{\end{eqnarray}}
\def\beq{\begin{equation}}   \def\eeq{\end{equation}}

\def\beeq{\begin{eqnarray}} \def\eeeq{\end{eqnarray}}

\def\be{\begin{equation}}
\def\ee{\end{equation}}
\def\bea{\begin{eqnarray}}
\def\eea{\end{eqnarray}}

\begin{document}
\vspace*{3.8cm}
\title{Baryon and Antibaryon production at RHIC energies in 
the Dual Parton Model}

\author{A. Capella$^{\rm a)}$, C.A. Salgado$^{\rm b)}$, \underline{D. Sousa}$^{\rm c)}$ }

\address{$^{\rm a)}$ Laboratoire de Physique Th\'eorique\footnote{Unit\'e Mixte de
Recherche UMR n$^{\circ}$ 8627 - CNRS}
\\ Universit\'e de Paris XI, B\^atiment 210,
F-91405 Orsay Cedex, France \\
$^{\rm b)}$ Theory Division, CERN, CH-1211 Geneva 23, Switzerland \\
$^{\rm c)}$ ECT*, Trento, Italy}

\vspace*{-0.2cm}
\maketitle\abstracts{We compute the mid-rapidity densities of pions, kaons,
baryons and antibaryons in $Au$--$Au$ collisions at $\sqrt{s}$ = 130 GeV
in the Dual Parton Model supplemented with final state interactions, and
we present a comparison with available data.}

\vspace*{-0.8cm}
\section{Description of the model}

\noindent
The rapidity density of a given type of hadron $h$ produced in $AA$
collisions  at fixed impact parameter, is given by \cite{1r,2r}

\bea
\label{1e}
{dN^{AA\to h} \over dy}(y,b) &=& n_A(b) \left [
N_{h,\mu(b)}^{qq^{P}-q_{v}^{T}} (y) + N_{h,\mu(b)}^{q_{v}^P -
qq^T} (y) + (2k-2) \
N_{h,\mu(b)}^{q_s - \overline{q}_s}(y)\right ] \nonumber \\
&&+  \left ( n(b) - n_A(b) \right )  2k \
N_{h,\mu(b)}^{q_s-\overline{q}_s} (y) \ . \eea

\noindent
Here $n(b)$ is the average number of binary collisions and $n_A(b)$
is the average number of participant pairs  at fixed impact 
parameter $b$.
$P$ and $T$ denote projectile and target nuclei.
$k$ is the average number of inelastic collisions in $pp$ and $\mu(b)
= k \nu (b)$ with $\nu (b) = n(b)
/n_A(b)$ the average total number of collisions suffered by each
nucleon. At $\sqrt{s} = 130$~GeV we have $k
= 2$ \cite{2r}. 
The $N_{h,\mu (b)}(y)$ in eq. (\ref{1e}) are the rapidity
distributions of hadron $h$ in each individual
string. In DPM they are given by convolutions of momentum
distribution and fragmentation
functions \cite{3r}.

It was shown in \cite{2r} that eq. (\ref{1e}), supplemented with
shadowing corrections, leads to values of
charged multiplicities at mid-rapidities as a function of centrality
in agreement with data, both
at SPS and RHIC. Here we use the same shadowing corrections as in
ref. \cite{2r} \\

{\it Net Baryon Production ($\Delta B = B-\overline{B}$):}
In order to reproduce the stopping observed in $Pb$ $Pb$ collisions 
at SPS, we introduce a new 
mechanism (diquark breaking)
based on the transfer in rapidity of the baryon junction
\cite{4r,4br}. In an $AA$ collision, this
component gives the following
rapidity distribution of the two net baryons in a single $NN$
collision of $AA$ \cite{6r}

\beq
\label{4e}
\left ( {dN_{DB}^{\Delta B} \over dy} (y) \right )_{\nu(b)} =
C_{\nu(b)} \left [ Z_+^{1/2} (1 - Z_+)^{\nu(b) - 3/2}
+ Z_-^{1/2} (1 - Z_-)^{\nu(b) - 3/2} \right ] \eeq

\noi where $Z_{\pm} = \exp (\pm y - y_{max})$ and $\nu(b) = n(b)/n_A(b)$.
$C_{\nu(b)}$ is determined from the
normalization to two at each $b$.  \par

In order to get the relative densities of each baryon and antibaryon
species we use simple quark counting rules \cite{6r}. We denote the
strangeness suppression factor by $S/L$ (with $2L+ S = 1$). 
If the baryon is made out of three sea quarks (which is the case 
of pair production) 
the relative weights are $I_3 = 4L^3 : 4L^3 : 12L^2S : 3LS^2 :
3LS^2 : S^3$
for $p$, $n$,
$\Lambda + \Sigma$, $\Xi^0$, $\Xi^-$ and $\Omega$, respectively. The
various coefficients of $I_3$ are obtained from the power expansion
of $(2L +
S)^3$. 
For net baryon production there are two possibilities: one is to use
$I3$ which corresponds to the transfer of $SJ$ without quarks.The other 
possibility corresponds to the trasfer of the baryon junction plus
one valence quark. In this case the relevant weights are given by $I_2$, i.e.
from the various terms in the expansion of $(2L +  S)^2$. This second
possibility is favored by data and it is the considered in this work.
In order to take into account the decay of $\Sigma^*(1385)$
into $\Lambda \pi$, we redefine the relative rate of $\Lambda$'s and
$\Sigma$'s using the empirical rule $\Lambda = 0.6(\Sigma^+ +
\Sigma^-$) -- keeping, of course, the total
yield of $\Lambda$'s plus $\Sigma$'s unchanged. In this way the
normalization constants of all baryon species in
pair production are determined from one of them. This constant,
together with the relative normalization of
$K$ and $\pi$, are determined from the data for very peripheral
collisions. In the calculations we use $S =
0.1$ $(S/L = 0.22)$. \par

\vskip 0.15cm
{\it Final State Interactions:}
The hadronic densities
obtained above will be used as initial
conditions in the gain and loss differential equations which govern
final state interactions.

\beq
\label{6e}
\tau {d\rho_i \over d \tau} = \sum_{k\ell} \sigma_{k\ell} \ \rho_k \
\rho_{\ell} - \sum_k \sigma_{ik} \ \rho_i
\ \rho_k \ . \eeq

\noi The first term in the r.h.s. of (\ref{6e}) describes the
production (gain) of particles of type $i$
resulting from the interaction of particles $k$ and $\ell$. The
second term describes the loss of particles
of type $i$ due to its interaction with particles of type $k$. In eq.
(\ref{6e}) $\rho_i = dN_i/dy d^2s(y,b)$ are the particles
yields per unit rapidity and per unit of transverse area, at fixed
impact parameter. $\sigma_{k\ell}$ are the
corresponding cross-sections averaged over the momentum
distribution of the colliding particles. \par

The channels that have been taken into account in
our calculations are

\beq
\label{7e}
\pi N \stackrel{\rightarrow}{\leftarrow} K \Lambda (\Sigma)\ , \quad
\pi \Lambda (\Sigma )
\stackrel{\rightarrow}{\leftarrow} K \Xi \ , \quad \pi \Xi
\stackrel{\rightarrow}{\leftarrow} K \Omega  \eeq

\noi 
Of all possible
charge combinations in these reactions, we have only kept those
involving the annihilation of a light $q$-$\overline{q}$ pair and
production of an $s$-$\overline{s}$ in the
$s$-channel.
We have also taken into account the strangeness exchange reactions

\beq
\label{8e}
\pi \Lambda (\Sigma ) \stackrel{\rightarrow}{\leftarrow} K N\ , \quad \pi \Xi
\stackrel{\rightarrow}{\leftarrow} K \Lambda (\Sigma ) \ , \quad \pi
\Omega  \stackrel{\rightarrow}{\leftarrow}
K \Xi \ . \eeq

\noi as well as the channels corresponding to (\ref{7e}) and
(\ref{8e}) for antiparticles.

\section{Results}

\begin{figure}
\begin{center}
\epsfig{figure=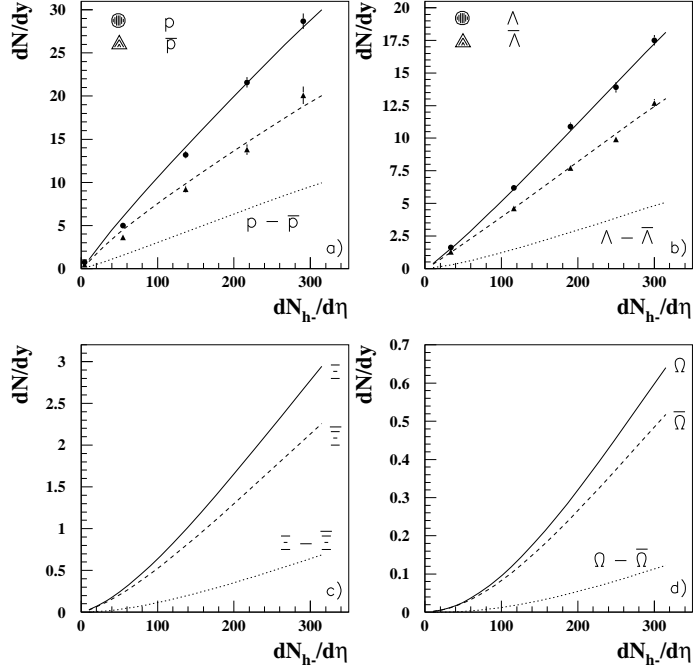,height=4.0in}
\end{center}
\caption{
(a) Calculated values of rapidity densities of
$p$ (solid line), $\overline{p}$ (dashed line), 
and $p - \overline{p}$ (dotted line) at mid rapidities, 
$|y^*| < 0.35$, are
plotted as a function of $dN^-/d\eta$, and
compared with PHENIX data \protect\cite{7r}~; 
(b) same for $\Lambda$ and
$\overline{\Lambda}$ compared to preliminary 
STAR data \protect\cite{9r}~;
(c) same for $\Xi^-$ and $\overline{\Xi}^+$~;
(d) same for $\Omega$ and
$\overline{\Omega}$~.}
\end{figure}

The calculations have been
performed in the interval $-0.35 < y^* < 0.35$. In Fig.~1a-1d we show
the rapidity densities of
$B$, $\overline{B}$ and $B - \overline{B}$ versus
$h^- = dN^-/d\eta = (1/1.17) dN/dy$ and compare them with available
data \cite{7r,8r,9r}. We see that, in first approximation, $p$,
$\overline{p}$,
$\Lambda$ and $\overline{\Lambda}$ scale with $h^-$. Quantitatively,
there is a slight decrease with centrality of
$p/h^-$ and $\overline{p}/h^-$ ratios, a slight increase of $\Lambda
/h^-$ and $\overline{\Lambda}/h^-$ and a much larger increase for
$\Xi$
($\overline{\Xi})/h^-$ and $\Omega$ ($\overline{\Omega})/h^-$.
In the DPM (before final state interaction) the
rapidity density of charged particle per participant increases with
centrality.
This increase is
larger for low centralities \cite{9r}.
This has an important effect on both the size and the
pattern of strangeness enhancement in our results.
It explains why the
departure from a linear increase of $\Xi$'s and $\Omega$'s (concave
shape) seen in Figs.~1c and 1d is also more pronounced for lower
centralities.

The ratios $\overline{B}/B$ have a mild decrease with
centrality of about
15~\% for all baryon species -- which is also seen in the data
\cite{10r}. Our values for $N^{ch}/N_{max}^{ch} = 1/2$ are~:
$\overline{p}/p = 0.69$, $\overline{\Lambda}/\Lambda = 0.72$,
$\overline{\Xi}/\Xi = 0.79$, $\Omega/\overline{\Omega} =
0.83$ to be compared with the
measured values \cite{8r}~: $$\overline{p}/p = 0.63 \pm 0.02 \pm
0.06 \quad , \quad
\overline{\Lambda}/\Lambda = 0.73 \pm 0.03 \quad , \quad
\overline{\Xi}/\Xi = 0.83 \pm 0.03 \pm 0.05 \ .$$

\begin{figure}
\begin{minipage}{40mm}
\centering\epsfig{file=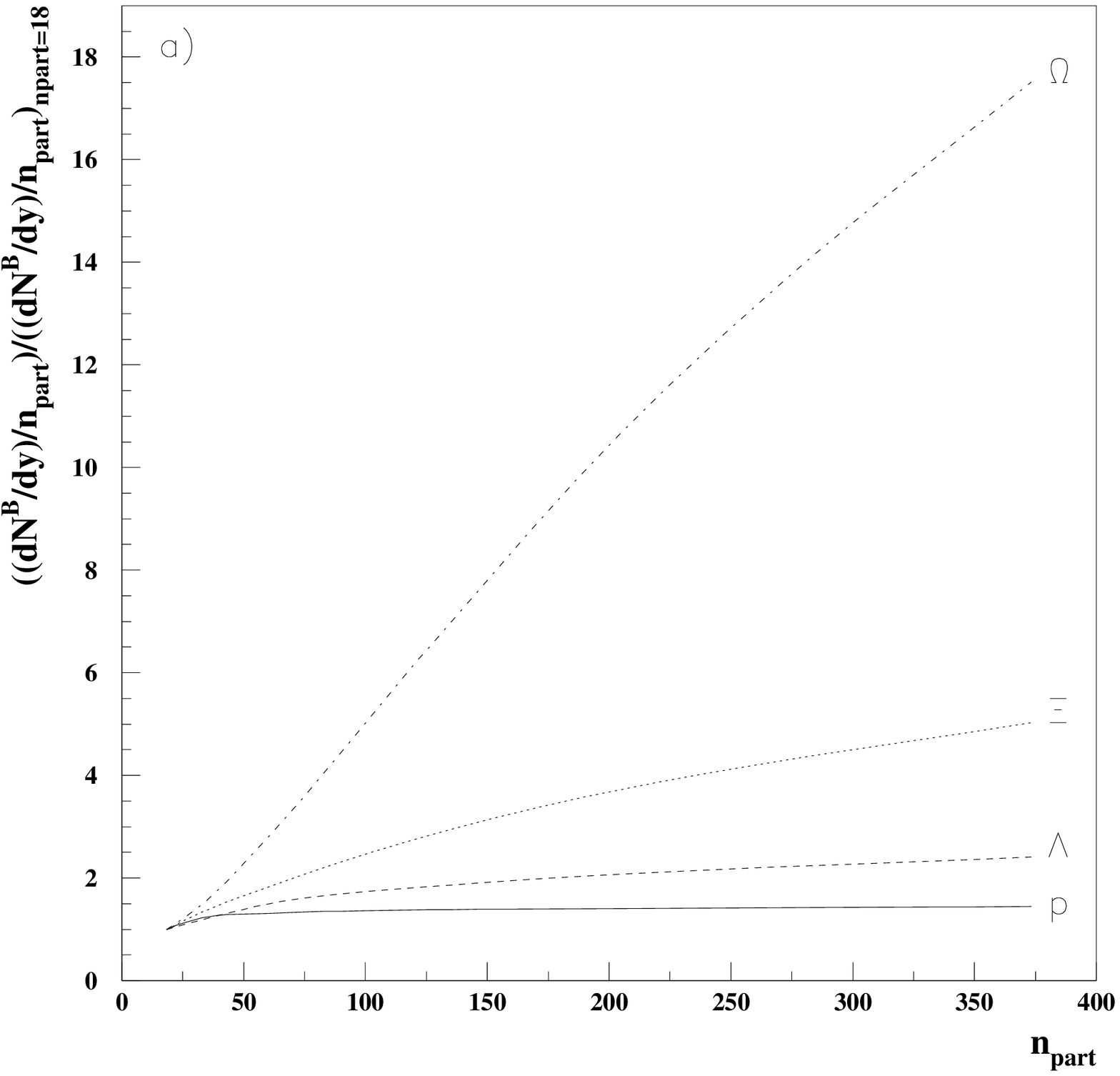,bbllx=0,bblly=30,bburx=540,bbury=620,height=3.0in}
\end{minipage}
\hspace{\fill}
\begin{minipage}{40mm}
\centering\epsfig{file=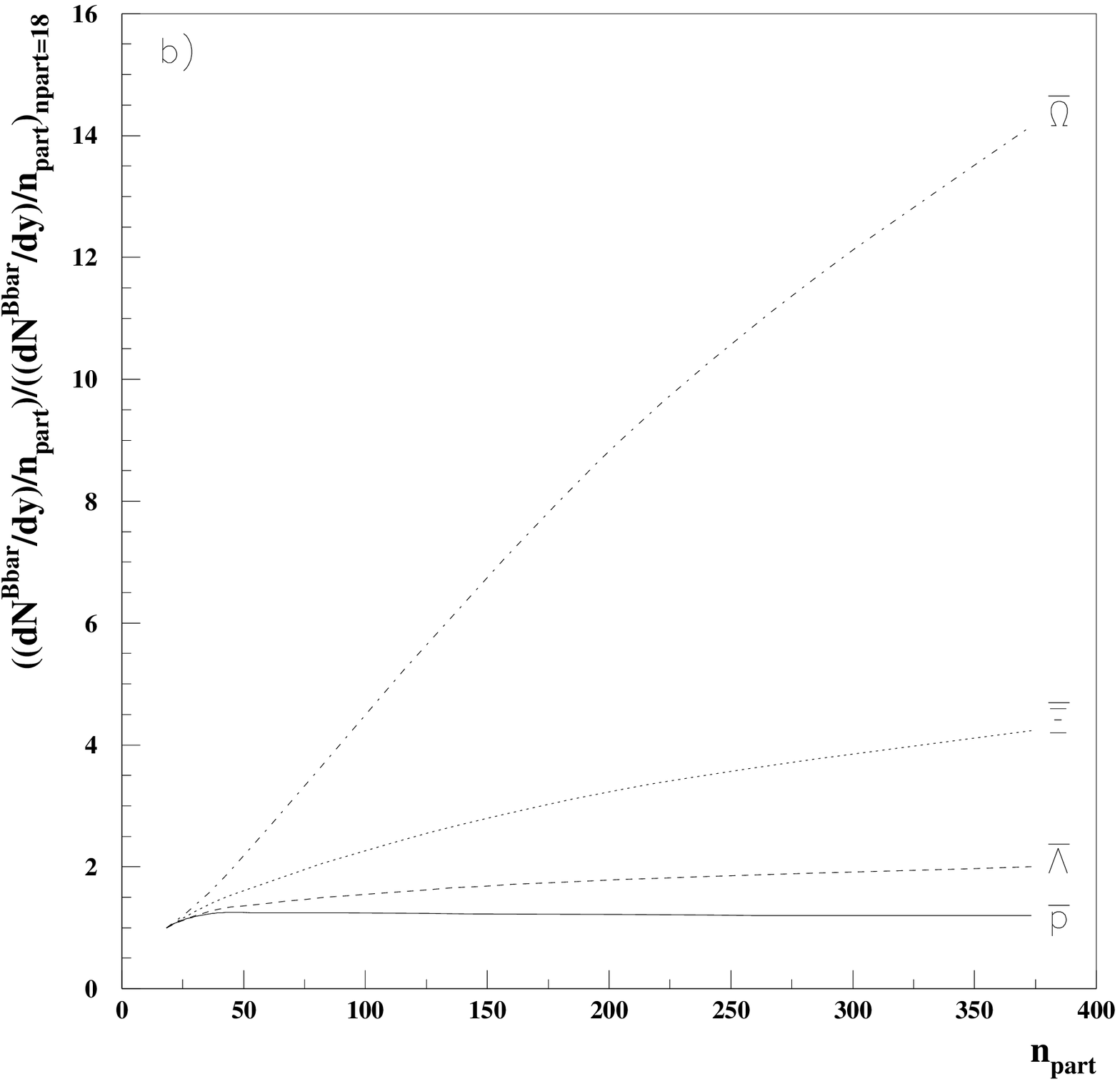,bbllx=280,bblly=30,bburx=840,bbury=620,height=3.0in}
\end{minipage}
\caption{Calculated values of the ratios $B/n_{part}$ (a)
and $\overline{B}/n_{part}$ (b), normalized
to the same ratio for peripheral collisions ($n_{part} = 18$), are plotted as a
function of $n_{part}$.}
\end{figure}

In Fig.~2a and 2b we plot the yields of $B$ and $\overline{B}$ per
participant normalized
to the same ratio for peripheral collisions versus $n_{part}$. The
enhancement of $B$ and $\overline{B}$ increases with the number of
strange quarks in
the baryon. This increase is comparable
to the one found at SPS between pA and
central PbPb collisions -- somewhat larger for antibaryons.
Before final state
interactions, all ratios $K/h^-$, $B/h^-$
and $\overline{B}/h^-$ decrease slightly with increasing centrality.
This effect is rather marginal at RHIC
energies and mid-rapidities. The final state interactions (\ref{7e}), (\ref{8e}) lead to a gain of
strange particle
yields.  The reason for this is the following. In the first direct
reaction (\ref{7e}) we have $\rho_{\pi} >
\rho_K$, $\rho_N > \rho_{\Lambda}$, $\rho_{\pi} \rho_N \gg \rho_K
\rho_{\Lambda}$. The same is true for all
direct reaction (\ref{7e}). In view of that, the effect of the
inverse reactions (\ref{7e}) is small. On the contrary, in
all reactions (\ref{8e}), the product of densities in the initial and
final state are comparable and the
direct and inverse reactions tend to compensate with each other.
Baryons with the largest strange quark
content, which find themselves at the end of the chain of direct
reactions (\ref{7e}) and have the smallest
yield before final state interaction, have the largest enhancement.
Moreover, the gain in the yield of
strange baryons is larger than the one of antibaryons since $\rho_B >
\rho_{\overline{B}}$. Furthermore, the
enhancement of all baryon species increases with centrality, since
the gain, resulting from the first term in
eq. (\ref{6e}), contains a product of densities and thus, increases
quadratically with increasing centrality.\par

\vspace*{-0.2cm}

\section*{Acknowledgments}
It is a pleasure to thank the organizers for a nice and
stimulating meeting. C.A.S. is supported by a Marie Curie
Fellowship of the European Community program TMR (Training and Mobility of
Researchers), under the contract number HPMF-CT-2000-01025.\\

\end{document}